**Néel-order-dependent transverse transport in noncoplanar antiferromagnet $MnTe_2$**

Qi Feng[1,#], Yilin Han[1,#], Yongkai Li[1], Yuqing Hu[1], Mo Tian[1], Qiuli Li[1], Huimin Peng[1], Jinrui Zhong[1], Zhiwei Wang[1,2,*], Zhi-Ming Yu[1,*], Junxi Duan[1,*], Yugui Yao[1,2,*]

[1]Key Laboratory of Advanced Optoelectronic Quantum Architecture and Measurement (MOE), School of Physics, Beijing Institute of Technology, Beijing 100086, China

[2]International Center for Quantum Materials, Beijing Institute of Technology, Zhuhai, 519000, China

# These authors contribute equally

* Corresponding authors

junxi.duan@bit.edu.cn

zhiweiwang@bit.edu.cn

zhiming_yu@bit.edu.cn

ygyao@bit.edu.cn

**Antiferromagnets hold appealing potential in next-generation spintronic devices with higher frequency and scalability, thanks to their alternating spin orientations that cancel out net magnetization. However, the lack of a nonzero magnetization makes the detection of the magnetic configuration of antiferromagnet difficult, hampering the applications of antiferromagnets. Here, we report a new transverse transport effect in noncoplanar antiferromagnet $MnTe_2$. This effect is antisymmetric in both magnetic field and Néel order, but symmetric in its two indices. It can be understood in terms of the contribution induced by both magnetic field and geometric quantities, as confirmed by our theoretical calculations. Our discovery of a new Néel-order-dependent transverse transport effect provides opportunities to the advancing antiferromagnetic spintronics.**

Spintronics, a fast-developing field emerges at the intersection of condensed matter physics and electronics, exploits the spin of electrons to encode and process information, transcending the limitations of conventional charge-based electronics [1-4]. It promises ultra-low-power and high-speed devices with applications ranging from magnetic memory to quantum computing [5,6]. Previous spintronics studies focus on the ferromagnets, especially in spin valves [7-9] and magnetic tunnel junctions [10-13], enabling applications in hard drives and magnetoresistive random-access memory [14-17]. However, ferromagnets face challenges, including susceptibility to external magnetic fields and energy losses from stray fields, limiting their scalability [18-21]. In contrast, antiferromagnets (AFM), with their alternating spin orientations that cancel out net magnetization, offer unique advantages including immunity to external magnetic perturbations, absence of stray fields, and ultrafast spin dynamics [22-25]. Recent breakthroughs in manipulating antiferromagnetic order via electric currents or optical pulses have positioned them as revolutionary candidates for next-generation spintronic devices, reigniting interest in their fundamental physics and practical potential [26-31].

However, the development of the antiferromagnetic spintronics is slow. The main reason is that without a net magnetization, the detection of the magnetic configuration of antiferromagnets, i.e. the Néel order characterized by the Néel vector $\boldsymbol{n}$, is notoriously difficult. Although recent progresses show that anomalous Hall effect (AHE) is allowed in altermagnets and anomalous-Hall antiferromagnets, which is closely related to the magnetic order, strong symmetry restrictions of the antiferromagnet hosting AHE limit the family of candidate materials [32-35]. Under these circumstances, a natural question arises: whether there are new Néel-order-dependent transport phenomena beyond the AHE?

In this work, we report the observation of a new Néel-order-dependent transverse transport effect in noncoplanar antiferromagnet $MnTe_2$. Through comparing the results measured after zero-field cooling (ZFC) and field-cooling (FC) under different magnetic fields, we have determined that the transverse response of $MnTe_2$ consists of the ordinary Hall effect (OHE) and an unconventional transverse transport effect. This prominent unconventional effect is identified to be antisymmetric in both magnetic field ($B$) and $\boldsymbol{n}$. It can be understood in terms of the contribution induced by both magnetic field and two geometric quantities: Berry curvature and orbital moment. This contribution is time-reversal antisymmetric but symmetric in its two indices, i.e. $\sigma_{xy} = \sigma_{yx}$, indicating a transverse transport rather than a Hall effect. Specifically, we find that switching the Néel order of $MnTe_2$ does not change its electronic bands; however, it flips the sign of the Berry curvature and the orbital moment, leading to the reversal of the unconventional transverse effect. Our study not only deepens the understanding of transport in antiferromagnets but also provides opportunities to the advancing antiferromagnetic spintronics.

**Field-cooling dependent transverse resistance**

$MnTe_2$ is a three-dimensional noncoplanar antiferromagnet belonging to magnetic space group 205.33, which includes spatial inversion symmetry $P$, threefold rotation symmetry $C_{3,111}$, and two twofold screw-rotation symmetry, $\{C_{2z}|1/2,0,1/2\}$ and $\{C_{2y}|0,1/2,1/2\}$ [36]. Its unit cell comprises four Mn atoms and eight Te atoms, with each Mn atom octahedrally coordinated by six equivalent Te atoms. The localized magnetic moments reside primarily on the four Mn atoms, labeled as $\mathbf{S}_1$, $\mathbf{S}_2$, $\mathbf{S}_3$, and $\mathbf{S}_4$ in Fig. 1(a). These moments can all point inward or outward, forming "all-in-all-out" AFM magnetic configurations [37,38], as illustrated in Figs. 1(a) and 1(b). The two Néel orders are degenerate in energy and respectively termed as $\alpha$-phase and $\beta$-phase here, which are connected by time-reversal symmetry *T*. It is interesting to note that only one of the four local moments are independent. For example, $\mathbf{S}_2$, $\mathbf{S}_3$, and $\mathbf{S}_4$ are connected by $C_{3,111}$, and $\mathbf{S}_1$ is related to $\mathbf{S}_2$ by $\{C_{2z}|1/2,0,1/2\}$. Thus, without loss of generality, we can define the Néel vector of $MnTe_2$ as $\boldsymbol{n}=\mathbf{S}_1$.

Our $MnTe_2$ crystals were grown using chemical vapor transport method with iodine as the transport agent, in a temperature gradient from 520 °C to 590 °C. Single crystals with a lateral dimension of 2-3 mm were polished by sand papers to get a thin flake with a thickness about hundreds of microns. Crystal structure and chemical composition were characterized by X-ray diffraction and energy-dispersive spectroscopy. Magnetization properties were measured by vibrating sample magnetometer. Figure 1(c) presents the ZFC temperature dependence of magnetization in sample N4 with $H \parallel (001)$, which shows a typical antiferromagnetic behavior with a Néel temperature, $T_N$, of 88.91 K. For transport measurements, electrodes with Hall-bar geometry were made on the samples by gold wires pasted with silver paste. Longitudinal and transverse resistance, $R_{xx}$ and $R_{yx}$, were measured with four-probe lock-in method and extracted by standard symmetrizing and anti-symmetrizing procedures, respectively [39]. We have measured three samples and observed consistent results. Results from sample N1 are discussed in the main text, and results from N2 and N3 can be found in the supplementary [39]. Figure 1(d) plots the temperature dependence of the resistance of N1 with current applied along the *x*-axis. Similar to prior studies, the resistance exhibits a semiconductor behavior. A clear resistance kink is observed around $T_N = 82\,K$, corresponding to the paramagnetic-to-antiferromagnetic phase transition of $MnTe_2$.

We then investigate the transverse magneto-transport properties of N1. The sample is firstly zero-field cooled from 90 K to 20 K. After being stabilized at 20 K, a perpendicular magnetic field (along the *z* axis) is applied upon the sample, and field dependence of $R_{xx}$ and $R_{yx}$ are extracted. The field dependence of $R_{yx}$ is shown in Fig. 2(a) (black line), exhibiting a positive slope. Importantly, $R_{yx}$ measured with field being swept in both positive and negative directions shows no sign of hysteresis [39]. This is consistent with the linear ZFC field dependence of magnetization shown in the inset of Fig. 1(c) (black line), which indicates that there is no ferromagnetism in $MnTe_2$, in agreement with prior studies [36]. We note that there is slope change in $R_{yx}$ at high field. However, considering the linear field dependence of magnetization, it cannot be interpreted by AHE but the existence of multiple carrier pockets on the Fermi surface, which is supported by the band structure shown in the Discussion section [53].

Interestingly, $R_{yx}$ of $MnTe_2$ shows intriguing behavior after field cooling. In Fig. 2(a), the red solid and dash lines show the $R_{yx}$ vs. $B$ curves measured at 20 K after the sample being field cooled from 90 K to 20 K under an initial magnetic field (IB) of 10 T and -10 T, respectively. The details of the measurements can be found in Supplemental Material [39]. Obviously, these two curves deviate severely from the ZFC one. Their slope increases significantly in magnitude. The slope of the curve under IB=-10 T (red dash line) even reverses its sign from positive to negative. We should mention that after field cooling procedure, $R_{yx}$ measured with field being swept in both positive and negative directions does not shown hysteresis either [39]. Meanwhile, we have measured the field dependence of magnetization in sample N4 under the same conditions as those in the transport measurement, i.e. with the sample being field cooled under 9 T and -9 T, as shown by red and blue lines in the inset of Fig. 1(c), respectively [39]. Both lines almost coincide with the ZFC one (black line in the inset of Fig. 1(c)). All these features indicate that FC procedure does not change the antiferromagnetic nature of the sample. There is no net magnetization or ferromagnetism in the sample after FC procedure either, and thus the intriguing transverse resistance behaviors cannot be from AHE.

Conventionally, under a perpendicular magnetic field, $R_{yx}$ is the Hall effect consisting of OHE and AHE. Since the AHE has been ruled out, we should check the possibilities from the OHE. The OHE, denoted as $R_{yx}^{O}$, arises from the Lorentz force felt by the charge carriers under magnetic field, which depends on the type and density of the carrier and the thickness of the sample, i.e. $R_{yx}^{O}(B) = \frac{B}{nt}$, where $n$ is the carrier density with its sign indicating the carrier type, and $t$ the thickness of sample. Hence, the observed change in $R_{yx}$ after FC procedure under different IBs might come from the change of carrier type or density caused by FC-induced reconstruction of the Fermi surface. To test this scenario, we investigate the thermoelectric properties of the sample. According to the generalized Mott formula [54], the Seebeck voltage is highly sensitive to the Fermi surface. If there is any field-induced variation in the Fermi surface, the Seebeck voltage after FC under different IBs will show noticeable variation. We perform the Seebeck voltage measurement with the setup illustrated in the inset of Fig. 1(d). After FC from 90 K to 20 K under 10 T or -10 T, the field is tuned back to zero, and then the Seebeck voltage is recorded. The measured zero-field Seebeck voltage under IB=10 T is $-4.23 \pm 0.2\ \mu V$, while it is $-4.09 \pm 0.15\ \mu V$ under IB=-10 T. Within the experimental error, the two values can be considered identical, which indicates that the Fermi surface remain unaffected after FC. Meanwhile, the negative sign of both values cannot interpret the sign change of the slope of the transverse resistance. Therefore, possible origins of the FC dependent $R_{yx}$ from the OHE can also be ruled out.

**Decompose the transverse resistance**

From the above results, there should be an unconventional component sensitive to IB in the observed $R_{yx}$, i.e. the measured $R_{yx}$ consists of the OHE and an unconventional component. Since the unconventional component has been detected in samples with different crystallographic orientations, carrier types, and carrier densities, it is an intrinsic effect in $MnTe_2$. As aforementioned, the OHE, i.e. $R^{O}_{yx}(B)$, is IB independent. On the contrary, the unconventional one is a function of both B and IB, denoted as $R^{M}_{yx}(IB, B)$. Hence, the field dependence of $R_{yx}$ under IB=10 T and -10 T can be written as $R_{yx}(IB = 10T, B) = R^{M}_{yx}(IB = 10T, B) + R^{O}_{yx}(B)$ and $R_{yx}(IB = -10T, B) = R^{M}_{yx}(IB = -10T, B) + R^{O}_{yx}(B)$, respectively. Considering that the sign of $R_{yx}$ reverses when IB is changed from 10 T to -10 T, it is reasonable to assume that $R^{M}_{yx}(IB, B)$ is IB antisymmetric, i.e. $R^{M}_{yx}(IB = -10T, B) = -R^{M}_{yx}(IB = 10T, B)$.

As a result, the OHE can be extracted from

$$R^{O}_{yx}(B) = \frac{R_{yx}(IB,B)+R_{yx}(-IB,B)}{2} \qquad (1).$$

The red circular dots in Fig. 2(b) plot the extracted OHE at 20 K. It is surprising to note that this extracted OHE is roughly identical to the ZFC $R_{yx}$ (black square dots in Fig. 2(b)). Such an equivalence means that $R_{yx}^M$ is zero when the sample is zero-field cooled from temperature above $T_N$. Considering that $R_{yx}^M$ is sensitive to IB and $MnTe_2$ has two degenerate Néel orders, it hints that $R_{yx}^M$ is determined by the Néel order of $MnTe_2$. In general, in a ZFC bulk antiferromagnet, domains with different Néel orders coexist in the system below $T_N$. On average, the bulk sample shows no net magnetic feature, i.e. the total Néel vector is zero. Correspondingly, the unconventional component from these domains cancel out each other, and thus only OHE remains. On the other hand, when the bulk antiferromagnet is cooled from temperature above $T_N$ under IB, domains with a certain Néel order might have a lower energy, and thus a higher probability to exist. As a result, $R_{yx}^M$ is dominated by the majority of domains which is determined by IB. We note that such a domain selection by FC procedure has been identified by direct domain characterization in several antiferromagnetic systems including $Cr_2O_3$ [55], $Cd_2Os_2O_7$ [56], and MnTe [30].

In addition, we can extract $R_{yx}^M$ from the measured $R_{yx}$ that

$$R_{yx}^M(IB,B) = \frac{R_{yx}(IB,B) - R_{yx}(-IB,B)}{2} \qquad (2).$$

Figure 2c plots the $R_{yx}^M(IB = 10T, B)$ (black solid line) and $R_{yx}^M(IB = -10T, B)$ (black dash line). Both lines linearly depend on $B$, which suggests that the magnetic configuration is robust against the field up to 10 T at 20 K. We further examined the dependence of $R_{yx}^M$ on the two indices, i.e. $x$ and $y$. To measure $R_{xy}$, the current is injected along $y$ direction and the transverse voltage is measured in $x$ direction; $R_{xy}^O$ and $R_{xy}^M$ are extracted from $R_{xy}$ with the same procedure formulated in Eq. (1) and (2). As expected, $R_{xy}^O$ (blue diamond dots in Fig. 2(b)) has similar magnitude as $R_{yx}^O$ but reverses it sign, which confirms the Hall nature of $R_{xy}^O$ and verifies the validity of Eq. (1) and (2). Interestingly, as shown in Fig. 2(c), $R_{xy}^M$ (red solid and dashed lines for IB=10 T and -10 T, respectively) is almost identical to $R_{yx}^M$ (black lines) under the same IB. When there are two time-reversal-symmetry breaking components in the system, the Onsager reciprocal relations are kept validated provided that both components are reversed. Since in our case that the $R_{xy}^M$ is controlled by IB, the Onsager reciprocal relation of $R_{xy}^M$ can be written as $R_{xy}^M(IB,B) = R_{yx}^M(-IB,-B)$. Then, from the result that $R_{xy}^M(IB,B) = R_{yx}^M(IB,B)$, we can derive that

$$R_{yx}^{M}(IB,B)=R_{xy}^{M}(IB,B)=-R_{xy}^{M}(IB,-B)=-R_{yx}^{M}(-IB,B) \quad (3)$$

which verifies the validity of the assumption that $R_{yx}^{M}$ is IB antisymmetric.

**The unconventional component $R_{yx}^{M}$**

From these consistent results, we have identified that $R_{yx}$ in $MnTe_2$ consists of the OHE and an unconventional component. The unconventional component is hinted to be determined by the Néel order of $MnTe_2$. To further verify this connection, we systematically investigated the dependence of $R_{yx}^{M}$ on the magnitude of IB, the starting FC temperature, denoted as the initial FC temperature (IFT), and the measuring temperature (see End Matter).

First, we fix the IFT to 90 K but vary the magnitude of IB, i.e. the sample is cooled from 90 K to 20 K under different IBs. The $R_{yx}$ measured at 20 K are decomposed into $R_{yx}^{O}$ and $R_{yx}^{M}$ according to Eq. (1) and (2). In Fig. 3(a), we plot the $R_{yx}$ under different IBs. One can see that the $R_{yx}$ curves under IBs with the same sign almost collapse into one line except that curves under near-zero IBs deviate slightly. For clarity, we plot the IB dependence of the slope of $R_{yx}^{M}$ vs. B curves in Fig. 3(b). The magnitude of the slope rapidly increases with increasing IB and saturates above 3 T.

Next, we fix the magnitude of IB to 10 T but vary the IFT, i.e. the sample is cooled from different IFTs to 20 K under 10 T or -10 T. Here, we note that to get rid of the influence from the last round of measurement, the sample is firstly zero-field cooled from 90 K to IFT, and then cooled from IFT to 20 K under nonzero IB [39]. In Fig. 3(c), we plot the $R_{yx}$ vs. B curves with different IFTs. One can see that as IFT decreases, these curves become flatter. Most strikingly, the $R_{yx}$ vs. B curve under IB=-10T even reverse its slope when IFT=40 K. To see the effect of IFT on $R_{yx}^{M}$ clearly, we plot the extracted $R_{yx}^{M}$ with IB=10 T under different IFTs in Fig. 3(d). As expected, $R_{yx}^{M}$ decreases towards zero with decreasing IFT.

Together with the temperature dependence of $R_{yx}^{M}$ presented in Fig. 5 (see End Matter), the connection between $R_{yx}^{M}$ and the Néel order of $MnTe_2$ is undoubtedly unveiled. As aforementioned, $MnTe_2$ has two Néel orders characterized by opposite Néel vectors which are degenerate in energy without external magnetic field. Hence, when the bulk sample is zero-field cooled below $T_N$, domains with the two orders coexist with same weight. However, when a magnetic field is applied, domains with one of the two orders would have a lower energy. Consequently, as the bulk sample is cooled under IB, domains with the energy preferred Néel order becomes the majority below $T_N$. Its weight increases with the increasing magnitude of IB and then saturates above some critical IB. This is exactly the dependence of $R_{yx}^M$ on IB shown in Fig. 3(b). On the other hand, although the two Néel orders have identical energy without magnetic field, it is not energy free to switch a domain from one order into the other below $T_N$. There is an energy barrier between the two orders, which increases with decreasing temperature. As a result, when a magnetic field is applied on the zero-field-cooled sample at IFT, which is below $T_N$, only part of the domains with the higher-energy order would be switched to the lower-energy one. This switching effect becomes weaker when the energy barrier increases with decreasing temperature. It agrees well with that $R_{yx}^M$ becomes zero with decreasing IFT shown in Fig. 3(d). In addition, under fixed IB and IFT, the weight of the domains with the lower-energy order decreases upon increasing the measuring temperature due to thermal excitation and decreasing energy barrier. Therefore, $R_{yx}^M$ should decrease with increasing measuring temperature and disappear above $T_N$, which is exactly the result shown in Fig. 5(d) in End Matter. All these features distinguish $R_{yx}^M$ from previously known transverse transport phenomena, such as anisotropic magnetoresistance, magnetochiral anisotropy, quantum linear magnetoresistance, and possible contributions induced by domain inhomogeneity (see Sec. 6 in Supplemental Material [39]).

**Discussions**

Having established the connection between $R_{yx}^M$ and the Néel order, we will explore a possible mechanism of this new effect. According to the semiclassical theory within the relaxation time approximation [57, 58], the transverse conductivity that is in direct proportion to the magnetic field $B_z$ can be expressed as $\sigma_{yx} = \sigma_{yx}^{even} + \sigma_{yx}^{odd}$, with

$$\sigma_{yx}^{\mathrm{even}} = e^3\tau^2 \int [d\boldsymbol{k}] \frac{\partial f_k^0}{\partial \varepsilon} v_y(\boldsymbol{v} \times \boldsymbol{B}) \cdot \frac{\partial v_x}{\partial \boldsymbol{k}},$$

$$\sigma_{yx}^{\mathrm{odd}} = e^2\tau B_z \int [d\boldsymbol{k}] \frac{\partial f_k^0}{\partial \varepsilon} \left[-e\Omega_{\boldsymbol{k}z} v_x v_y - v_y \partial_{k_x} m_z - v_x \partial_{k_y} m_z\right].$$

The first term $\sigma_{yx}^{even}$ is induced by the Lorentz force, and gives rise to the OHE. Since both $\alpha$

and $\beta$ phases of $MnTe_2$ possess inversion symmetry $P$, the band structures of each phase have $\varepsilon_{\alpha(\beta)}(\boldsymbol{k}) = \varepsilon_{\alpha(\beta)}(-\boldsymbol{k})$. Moreover, because the two phases are connected by $T$, one also has $\varepsilon_{\alpha(\beta)}(\boldsymbol{k}) = \varepsilon_{\beta(\alpha)}(-\boldsymbol{k})$. Then, an interesting feature of $MnTe_2$ is obtained, i.e. the band structure of the two phases are identical $\varepsilon_{\alpha}(\boldsymbol{k}) = \varepsilon_{\beta}(\boldsymbol{k})$, as shown in Figs. 4(a) and 4(b). This means that $\sigma_{yx}^{even}$ is the same for both $\alpha$ and $\beta$ phases, and thus is an even function of the Néel vector of $MnTe_2$. In contrast, the second term $\sigma_{yx}^{odd}$ has topological origins, arising from the Berry curvature and the orbital momentum. Particularly, while $\sigma_{yx}^{odd}$ is antisymmetric in $B_z$, it is symmetric in its two indices, i.e. $\sigma_{yx}^{odd} = \sigma_{xy}^{odd}$, showing that it is a novel transverse transport rather than a Hall effect [39]. The Berry curvature and the orbital angular momentum are subject to same symmetry constraints, remaining invariant under $P$ while undergoing a sign reversal under $T$ [59]. Thus, we have $\Omega_{\alpha}(\boldsymbol{k}) = \Omega_{\alpha}(-\boldsymbol{k}) = -\Omega_{\beta}(\boldsymbol{k})$, and $m_{\alpha}(\boldsymbol{k}) = -m_{\beta}(\boldsymbol{k})$. To illustrate this directly, we plot the calculated distribution of the $z$ component of the Berry curvature for $\alpha$ and $\beta$ phases of $MnTe_2$ in a generic plane ($k_z$=-0.1 plane) of the Brillouin zone in Figs. 4(c) and 4(d) [39]. These numerical calculations are consistent with the symmetry analysis. Substituting the above identities into the expression of $\sigma_{yx}^{odd}$, it can be easily found that $\sigma_{yx}^{odd}$ has different sign for $\alpha$ and $\beta$ phases, leading to $\sigma_{yx}^{odd}(\boldsymbol{n}) = -\sigma_{yx}^{odd}(-\boldsymbol{n})$, and then is an odd function of the Néel vector of $MnTe_2$. These theoretical analyses agree well with our experimental results. However, our discussions do not rule out the contributions from other mechanisms, such as skew scattering and side-jump. It will be significant to further investigate the contributions of other mechanisms in the future.

In conclusion, we have discovered an unconventional transverse transport effect in noncoplanar antiferromagnet $MnTe_2$. It is identified to be determined by the Néel vector of $MnTe_2$ which can be controlled by external magnetic field during cooling. This novel Néel-order-dependent transverse transport effect could originate from the contribution induced by both magnetic field and Berry-curvature. Our work unveils a new effect for Néel order detection in antiferromagnets, which shed light on the spintronics applications of antiferromagnets.

## End Matter

### Temperature dependence of the unconventional component $R_{yx}^{M}$

We have measured the temperature dependence of the unconventional component $R_{yx}^{M}$ in sample N1. Here, we point out that $MnTe_2$ is not air stable, even they are stored in glovebox filled with nitrogen gas. Sample N1 has been stored in the glovebox for over one year since we finished the measurements presented in the main text. Therefore, we denote it as sample N1 (second time). To make good ohmic contact on the sample, we have slightly sanded the sample surface. From the R-T curve and ZFC OHE plotted in Fig. 5(a), we can see that the Néel temperature of the sample increases from 82 K to 87 K and the OHE under 10 T increases from $\sim$1.2 Ω to $\sim$2.5 Ω. These changes suggest that the sample has been doped during the storage and fabrication process that the chemical potential shifts accordingly. Nevertheless, the unconventional effect still exists, although its sign is reversed (this sign reversal will be discussed below). Considering that the Néel temperature of the sample increases to 87 K, we increase the IFT to 100 K to measure the temperature dependence of $R_{yx}^{M}$, i.e. the sample is zero-field cooled or field cooled under IB=+/-10 T from 100 K.

In Fig. 5(b), we have shown $R_{yx}$ measured at several typical temperatures between 20 K and 95 K after FC from 100 K under field with the solid lines and dashed lines standing for IB=10 T and -10 T, respectively. A full presentation of the temperature dependence of $R_{yx}$ and the details of the measurements can be found in Supplemental Material [39]. Meanwhile, the extracted field dependence of $R_{yx}^{M}$ under IB=10 T is presented in Fig. 5(c). To have a clearer view of the temperature dependence of $R_{yx}^{M}$, we have also plotted the temperature dependence of the low-field slope of $R_{yx}^{M}$ vs. B curves in Fig. 5(d) (because $R_{yx}^{M}$ is not linear to B at higher temperatures, which will be discussed below).

For these results, we can see several distinct features. (1) On the whole, the magnitude of $R_{yx}^{M}$ decreases with increasing temperature. It is notably nonzero at 80 K but becomes negligible at 95 K. This behavior strongly supports the connection between $R_{yx}^{M}$ and the antiferromagnetic transition, i.e. the Néel order. (2) $R_{yx}^{M}$ reverses its sign around 50 K. This sign reversal is consistent to the sign reversal at 20 K induced by sample change, considering that the sample change also affects the OHE. We can see that as the OHE at 10 T increases, the sign of $R_{yx}^{M}$ switches from positive to negative. Apparently, this sign reversal is related to the chemical potential shift. Taking that there are several carrier pockets on the Fermi surface, the measured $R_{yx}^{M}$ is the summation of the contributions from these carrier pockets. When the chemical potential is shifted, the weight of these pockets as well as their contributions change accordingly, which causes the reversal of the sign of $R_{yx}^{M}$. (3) We can see clear bending in the field dependence of $R_{yx}^{M}$ when temperature is above 40 K. The corresponding field where the bending happens decreases when the temperature is increased. In addition, the bending is somewhat abrupt at low temperatures, such as between 40 K and 55 K, but becomes quite smooth above 70 K. All these features suggest that it is related to field-induced flipping of the Néel order. Let us clarify the process. We assume that the sample is cooled under 10 T, and domains with Néel order $\boldsymbol{n}$ is preferred and the majority. When the magnetic field is swept from 10 T to -10 T, the energy of domains with $\boldsymbol{n}$ will increase while the one of domains with $-\boldsymbol{n}$ decreases. Once the field goes negative, the domains with $-\boldsymbol{n}$ will become preferred. At very low temperature, e.g. 20 K, the energy barrier between the two orders is high enough to prohibit the flip of the Néel order and no bending happens in $R_{yx}^{M}$. However, as the temperature increases, the barrier becomes lower that the flipping is allowed at some negative critical field. Apparently, the lower the barrier, the smaller this critical flipping field. Here, we want to point out that the flipping only happens under field with the direction opposite to IB, e.g. negative when the sample is cooled under 10 T. Since $R_{yx}^{M}$ is extracted from $R_{yx}$ measured under +/- IB with Eq. (2), the bending shows up on both positive and negative fields. We also note that no bending is seen in OHE, which is consistent with our interpretation that OHE is independent on the Néel order. When the temperature approaches the Néel temperature, the barrier is quite low and thermal excitation will broaden the flipping field, which smooths the bending of the line.

## References

[1] S. A. Wolf, D. D. Awschalom, R. A. Buhrman, J. M. Daughton, S. von Molnár, M. L. Roukes, A. Y. Chtchelkanova, and D. M. Treger, Spintronics: A Spin-Based Electronics Vision for the Future, Science **294**, 1488 (2001).

[2] B. Dieny *et al.*, Opportunities and challenges for spintronics in the microelectronics industry, Nat Electron **3**, 446 (2020).

[3] A. Hirohata, K. Yamada, Y. Nakatani, I.-L. Prejbeanu, B. Diény, P. Pirro, and B. Hillebrands, Review on spintronics: Principles and device applications, J Magn Magn Mater **509**, 166711 (2020).

[4] Z. Jia *et al.*, Spintronic Devices upon 2D Magnetic Materials and Heterojunctions, ACS Nano **19**, 9452 (2025).

[5] G.-M. Choi, O. Lee, S. Chung, W. Kim, T. Lee, B.-G. Park, and S. Yang, Spintronics and magnetic memory devices, Advances in Physics: X **10**, 2557918 (2025).

[6] Q. Shao, Z. Wang, Y. Zhou, S. Fukami, D. Querlioz, and L. O. Chua, Spintronic memristors for computing, npj Spintronics **3**, 16 (2025).

[7] B. Dieny, Giant magnetoresistance in spin-valve multilayers, J Magn Magn Mater **136**, 335 (1994).

[8] J. C. S. Kools, Exchange-biased spin-valves for magnetic storage, Ieee T Magn **32**, 3165 (1996).

[9] B. G. Park *et al.*, A spin-valve-like magnetoresistance of an antiferromagnet-based tunnel junction, Nat Mater **10**, 347 (2011).

[10] M. Julliere, Tunneling between ferromagnetic films, Phys Lett A **54**, 225 (1975).

[11] J. S. Moodera, L. R. Kinder, T. M. Wong, and R. Meservey, Large Magnetoresistance at Room-Temperature in Ferromagnetic Thin-Film Tunnel-Junctions, Phys Rev Lett **74**, 3273 (1995).

[12] S. Yuasa, T. Nagahama, A. Fukushima, Y. Suzuki, and K. Ando, Giant room-temperature magnetoresistance in single-crystal Fe/MgO/Fe magnetic tunnel junctions, Nat Mater **3**, 868 (2004).

[13] M. Piquemal-Banci, R. Galceran, M.-B. Martin, F. Godel, A. Anane, F. Petroff, B. Dlubak, and P. Seneor, 2D-MTJs: introducing 2D materials in magnetic tunnel junctions, Journal of Physics D: Applied Physics **50**, 203002 (2017).

[14] S. Bhatti, R. Sbiaa, A. Hirohata, H. Ohno, S. Fukami, and S. N. Piramanayagam, Spintronics based random access memory: a review, Materials Today **20**, 530 (2017).

[15] T. Nozaki, T. Yamamoto, S. Miwa, M. Tsujikawa, M. Shirai, S. Yuasa, and Y. Suzuki, Recent Progress in the Voltage-Controlled Magnetic Anisotropy Effect and the Challenges Faced in Developing Voltage-Torque MRAM, Micromachine **10**. 327 (2019).

[16] H. Yang *et al.*, Two-dimensional materials prospects for non-volatile

spintronic memories, Nature **606**, 663 (2022).

[17] V. D. Nguyen, S. Rao, K. Wostyn, and S. Couet, Recent progress in spin-orbit torque magnetic random-access memory, npj Spintronics **2**, 48 (2024).

[18] M. Jourdan *et al.*, Direct observation of half-metallicity in the Heusler compound $Co_2MnSi$, Nat Commun **5**, 3974 (2014).

[19] X. Q. Zhang *et al.*, Direct observation of high spin polarization in $Co_2FeAl$ thin films, Sci Rep-Uk **8**, 8074 (2018).

[20] C. Gong *et al.*, Discovery of intrinsic ferromagnetism in two-dimensional van der Waals crystals, Nature **546**, 265 (2017).

[21] Y. J. Deng *et al.*, Gate-tunable room-temperature ferromagnetism in two-dimensional $Fe_3GeTe_2$, Nature **563**, 94 (2018).

[22] T. Jungwirth, X. Marti, P. Wadley, and J. Wunderlich, Antiferromagnetic spintronics, Nat Nanotechnol **11**, 231 (2016).

[23] V. Baltz, A. Manchon, M. Tsoi, T. Moriyama, T. Ono, and Y. Tserkovnyak, Antiferromagnetic spintronics, Rev Mod Phys **90**, 015005 (2018).

[24] T. Jungwirth, J. Sinova, A. Manchon, X. Marti, J. Wunderlich, and C. Felser, The multiple directions of antiferromagnetic spintronics, Nat Phys **14**, 200 (2018).

[25] J. Han, R. Cheng, L. Liu, H. Ohno, and S. Fukami, Coherent antiferromagnetic spintronics, Nat Mater **22**, 684 (2023).

[26] Y.-W. Oh *et al.*, Field-free switching of perpendicular magnetization through spin–orbit torque in antiferromagnet/ferromagnet/oxide structures, Nat Nanotechnol **11**, 878 (2016).

[27] Y. Cheng, S. Yu, M. Zhu, J. Hwang, and F. Yang, Electrical Switching of Tristate Antiferromagnetic Neel Order in $\alpha$-$Fe_2O_3$ Epitaxial Films, Phys Rev Lett **124**, 027202 (2020).

[28] W. He *et al.*, Electrical switching of the perpendicular Néel order in a collinear antiferromagnet, Nat Electron **7**, 975 (2024).

[29] L. Huang *et al.*, Antiferromagnetic magnonic charge current generation via ultrafast optical excitation, Nat Commun **15**, 4270 (2024).

[30] O. J. Amin *et al.*, Nanoscale imaging and control of altermagnetism in MnTe, Nature **636**, 348 (2024).

[31] Z. Y. Zhou, X. K. Cheng, M. L. Hu, R. Y. Chu, H. Bai, L. Han, J. W. Liu, F. Pan, and C. Song, Manipulation of the altermagnetic order in CrSb via crystal symmetry, Nature **638**, 645 (2025).

[32] Z. X. Feng *et al.*, An anomalous Hall effect in altermagnetic ruthenium dioxide, Nat Electron **5**, 735 (2022).

[33] R. D. G. Betancourt *et al.*, Spontaneous Anomalous Hall Effect Arising from an Unconventional Compensated Magnetic Phase in a Semiconductor, Phys Rev Lett **130**, 036702 (2023).

[34] H. Reichlova *et al.*, Observation of a spontaneous anomalous Hall response in the $Mn_5Si_3$ d-wave altermagnet candidate, Nat Commun **15**, 4961 (2024).

[35] R. Takagi *et al.*, Spontaneous Hall effect induced by collinear

antiferromagnetic order at room temperature, Nat Mater **24**, 63 (2025).

[36] Y. P. Zhu *et al.*, Observation of plaid-like spin splitting in a noncoplanar antiferromagnet, Nature **626**, 523 (2024).

[37] N. J. Ghimire, A. S. Botana, J. S. Jiang, J. Zhang, Y. S. Chen, and J. F. Mitchell, Large anomalous Hall effect in the chiral-lattice antiferromagnet CoNb3S6, Nat Commun **9**, 3280 (2018).

[38] H. Takagi *et al.*, Spontaneous topological Hall effect induced by non-coplanar antiferromagnetic order in intercalated van der Waals materials, Nat Phys **19**, 961 (2023).

[39] See Supplemental Material at [URL] for Methods, Temperature dependence of $R_{yx}^{M}$, Data from samples N2 and N3, $R_{yx}$ measured with field swept in both positive and negative directions, Antisymmetrized longitudinal magneto-resistance $R_{xx}^{anti}$, Exclusion of different types of transverse transport effects, Distribution of the Berry curvature for $\alpha$ and $\beta$ phases of $MnTe_2$, and other supporting data images Fig. S1-S8, which includes Refs. [40-52].

[40] G. Kresse and J. Furthmüller, Efficient iterative schemes for ab initio total-energy calculations using a plane-wave basis set, Phys Rev B **54**, 11169 (1996).

[41] P. E. Blöchl, Projector augmented-wave method, Phys Rev B **50**, 17953 (1994).

[42] J. P. Perdew, K. Burke, and M. Ernzerhof, Generalized Gradient Approximation Made Simple, Phys Rev Lett **77**, 3865 (1996).

[43] J. P. Perdew, K. Burke, and M. Ernzerhof, Perdew, Burke, and Ernzerhof Reply, Phys Rev Lett **80**, 891 (1998).

[44] H. J. Monkhorst and J. D. Pack, Special points for Brillouin-zone integrations, Phys Rev B **13**, 5188 (1976).

[45] V. I. Anisimov, J. Zaanen, and O. K. Andersen, Band theory and Mott insulators: Hubbard U instead of Stoner I, Phys Rev B **44**, 943 (1991).

[46] A. A. Mostofi, J. R. Yates, G. Pizzi, Y.-S. Lee, I. Souza, D. Vanderbilt, and N. Marzari, An updated version of wannier90: A tool for obtaining maximally-localised Wannier functions, Computer Physics Communications **185**, 2309 (2014).

[47] Soumya Sankar *et al.*, Room temperature observation of the anomalous in-plane Hall effect in a Weyl ferromagnet, Nat Commun **17**, 423 (2025).

[48] T. Morimoto and N. Nagaosa, Chiral Anomaly and Giant Magnetochiral Anisotropy in Noncentrosymmetric Weyl Semimetals, Phys Rev Lett **117**, 146603 (2016).

[49] N. P. Ong and S. Liang, Experimental signatures of the chiral anomaly in Dirac–Weyl semimetals, Nat Rev Phys **3**, 394 (2021).

[50] G. L. J. A. Rikken, J. Fölling, and P. Wyder, Electrical Magnetochiral Anisotropy, Phys Rev Lett **87**, 236602 (2001).

[51] T. McGuire and R. Potter, Anisotropic magnetoresistance in ferromagnetic 3d alloys, Ieee T Magn **11**, 1018 (1975).

[52] A. A. Abrikosov, Quantum linear magnetoresistance, Europhysics Letters **49**,

789 (2000).

[53] N. Ashcroft and N. J. S. Mermin, Solid state physics; Thomson Learning Inc., 240 (1976).

[54] Y. M. Zuev, W. Chang, and P. Kim, Thermoelectric and Magnetothermoelectric Transport Measurements of Graphene, Phys Rev Lett **102**, 096807 (2009).

[55] P. J. Brown, J. B. Forsyth, and F. Tasset, A study of magnetoelectric domain formation in $Cr_2O_3$, Journal of Physics: Condensed Matter **10**, 663 (1998).

[56] S. Tardif, S. Takeshita, H. Ohsumi, J.-i. Yamaura, D. Okuyama, Z. Hiroi, M. Takata, and T.-h. Arima, All-In--All-Out Magnetic Domains: X-Ray Diffraction Imaging and Magnetic Field Control, Phys Rev Lett **114**, 147205 (2015).

[57] D. Ma, H. Jiang, H. Liu, and X. C. Xie, Planar Hall effect in tilted Weyl semimetals, Phys Rev B **99**, 115121 (2019).

[58] L. Li, J. Cao, C. X. Cui, Z. M. Yu, and Y. G. Yao, Planar Hall effect in topological Weyl and nodal-line semimetals, Phys Rev B **108**, 085120 (2023).

[59] D. Xiao, M. C. Chang, and Q. Niu, Berry phase effects on electronic properties, Rev Mod Phys **82**, 1959 (2010).

## Figures and captions

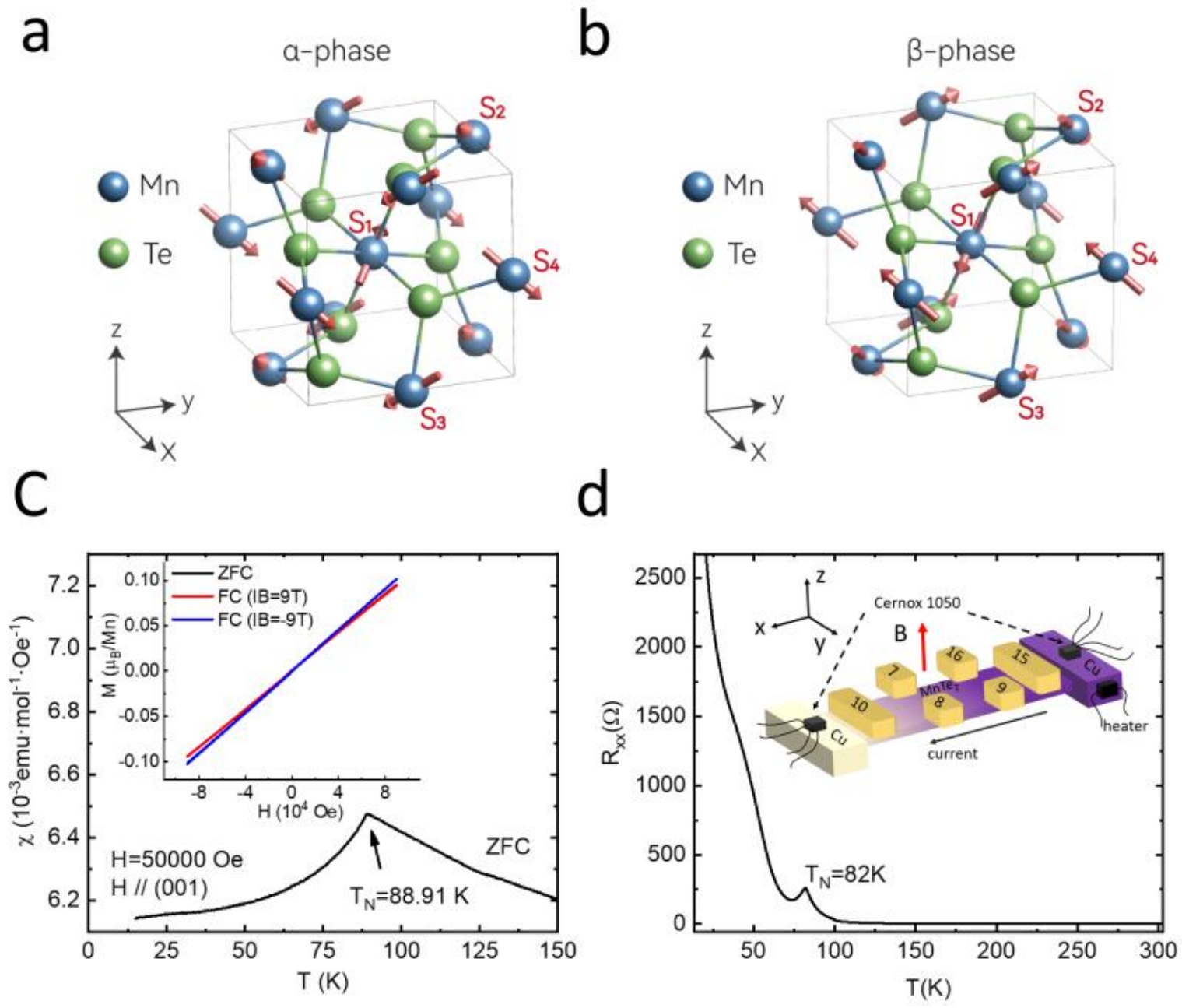

**Figure 1** Crystal structure of $MnTe_2$, magnetic properties of sample N4, and electrical transport properties of sample N1. (a) and (b) Crystal structures and magnetic configurations for $\alpha$-phase and $\beta$-phase $MnTe_2$, respectively. (c) Temperature dependence of magnetic susceptibility of sample N4, showing a typical antiferromagnetic behavior with a Néel temperature of 88.91 K. The inset shows the field dependence of magnetization of sample N4 measured at 20 K with the sample being zero-field cooled (ZFC) and field cooled (FC) under +/-9 T from 100 K. (d) Temperature-dependent resistance of sample N1, showing a Néel temperature at 82.0 K. The inset shows a schematic of the thermoelectric measurement setup and corresponding electrode configuration.

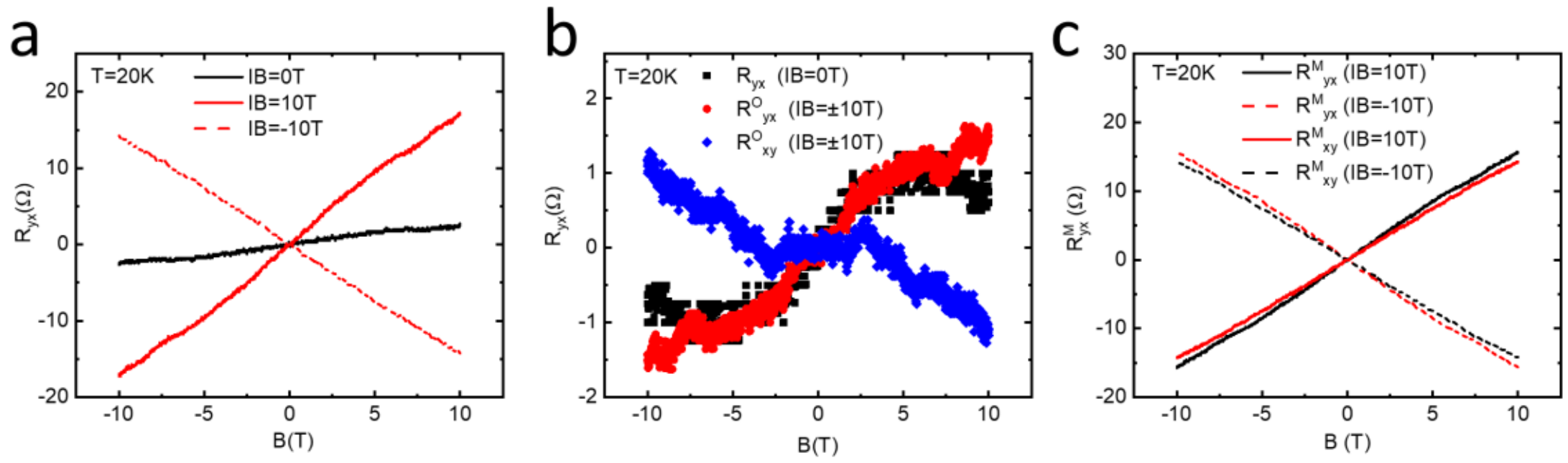

**Figure 2** Anomalous transverse resistance behavior in sample N1. (a) Transverse resistance $R_{yx}$ as a function of magnetic field measured after ZFC from 90 K to 20 K (black line) and after FC from 90 K to 20 K under 10T (red solid line) and -10T (red dashed line). (b) The ordinary Hall effect $R_{yx}^{O}$ (red circles), obtained from Eq. (1), is identical to the ZFC $R_{yx}$ (black squares). The ordinary Hall effect with interchanged indices $R_{xy}^{O}$ (blue diamonds) has the same magnitude as $R_{yx}^{O}$ but opposite sign. (c) Unconventional component $R_{yx}^{M}$ (black lines) and $R_{xy}^{M}$ (red lines) after field cooling from 90 K to 20 K under 10T (solid lines) and −10T (dashed lines).

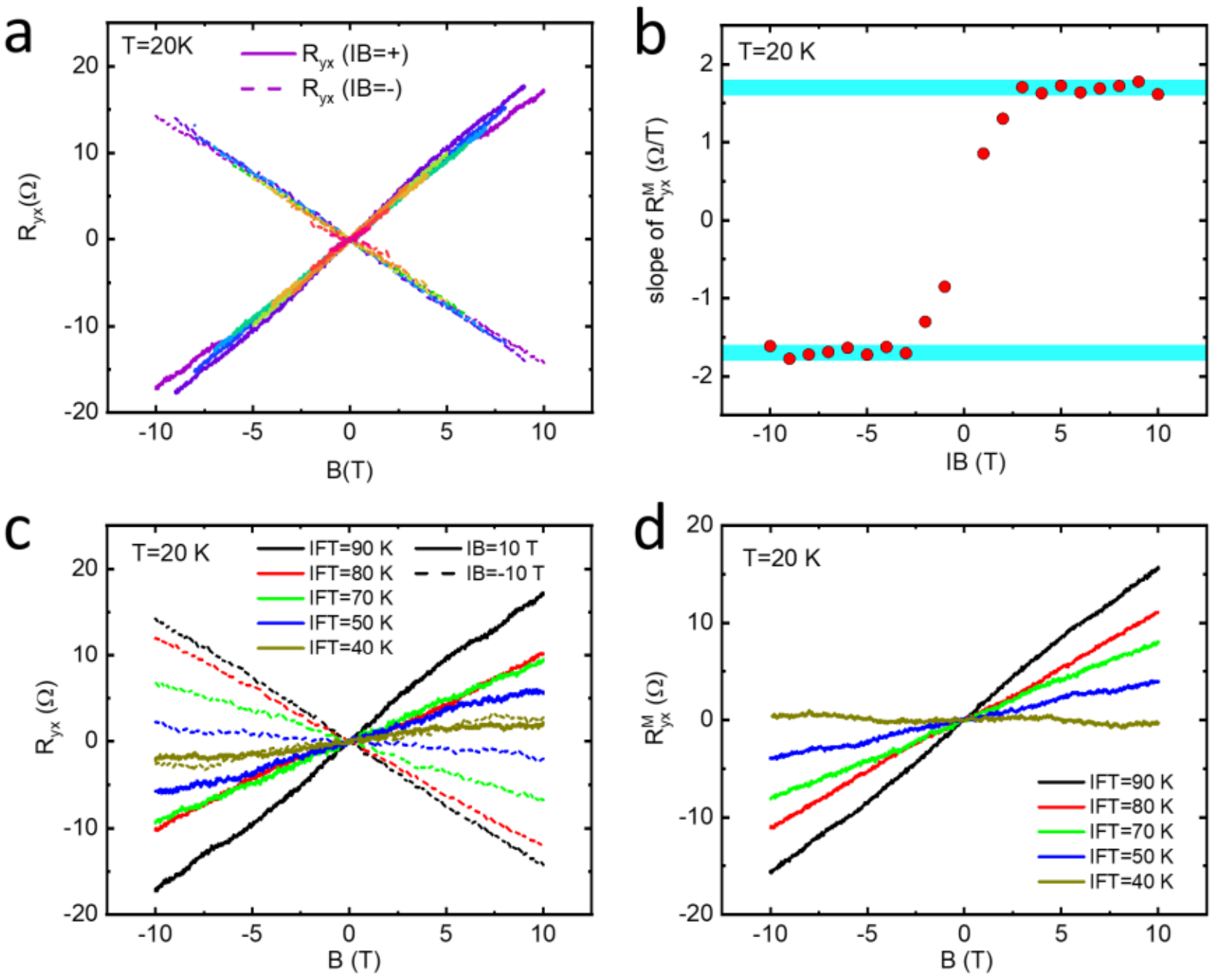

**Figure 3** Dependence of unconventional component on initial magnetic field (IB) and initial FC temperature (IFT) in sample N1. (a) Transverse resistance $R_{yx}$ as a function of magnetic field measured at 20 K after FC from 90 K under different IBs. Solid lines stand for positive IBs while dashed lines denote negative IBs. (b) Slope of $R_{yx}^{M}$ plotted as a function of IB, derived from the lines in (a). (c) Transverse resistance $R_{yx}$ measured at 20 K, after FC from different IFTs under a fixed IB of 10 T (solid lines) or −10 T (dashed lines). (d) $R_{yx}^{M}$ under IB=10 T with different IFTs extracted from $R_{yx}$ results in (c).

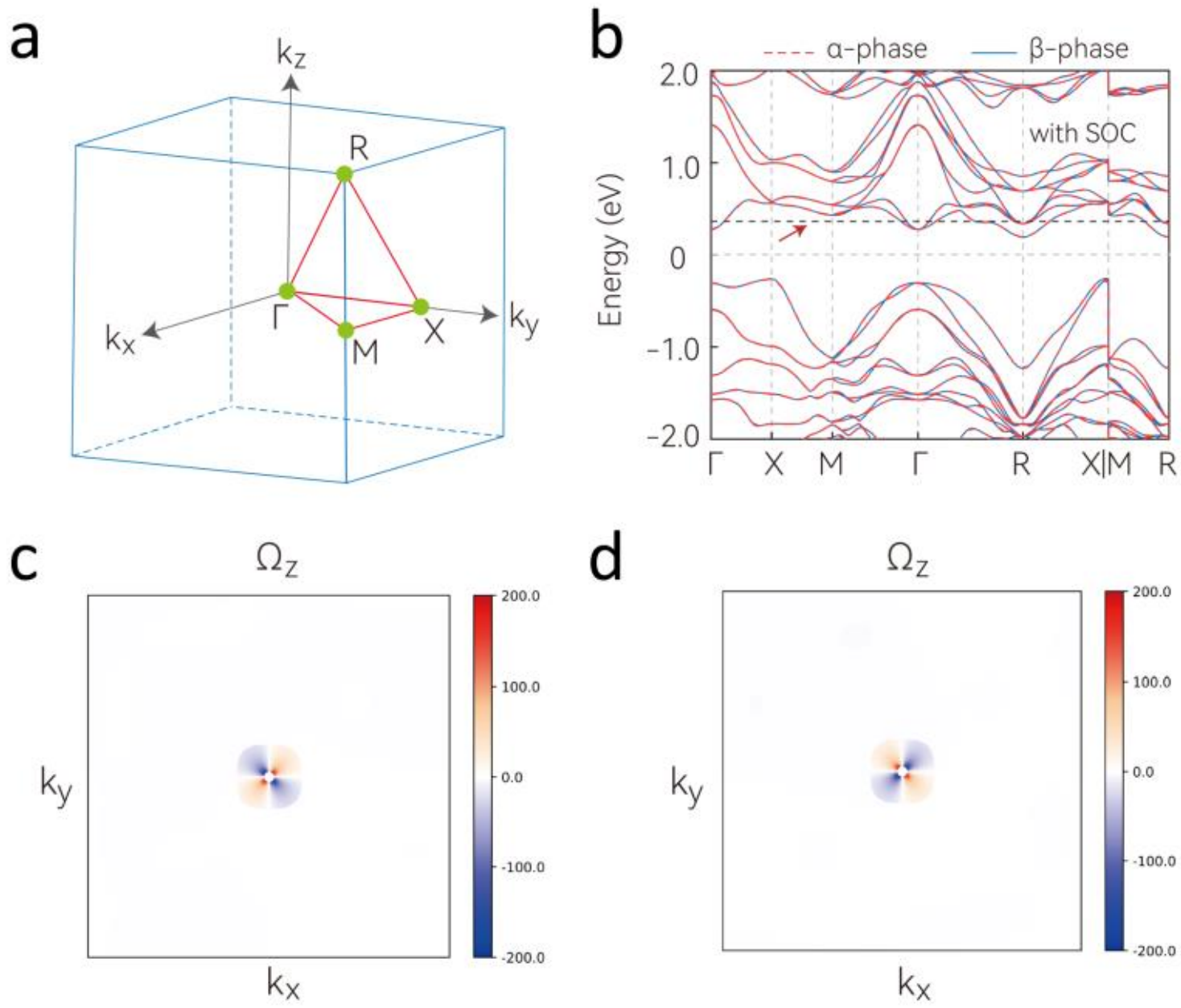

**Figure 4** Theoretical calculation of the electronic band structure and Berry curvature of $MnTe_2$. (a) The 3D Brillouin zone of $MnTe_2$. (b) The band structures of $\alpha$ and $\beta$ phases of $MnTe_2$ with spin-orbit coupling. (c-d) The calculated *z* component of the Berry curvature $\boldsymbol{\Omega}(\boldsymbol{k})$ in the $k_z$=-0.1 plane for (c) $\alpha$-phase and (d) $\beta$-phase $MnTe_2$, respectively. The Fermi energy used for calculating Berry curvature is 0.35 eV as indicated in (b).

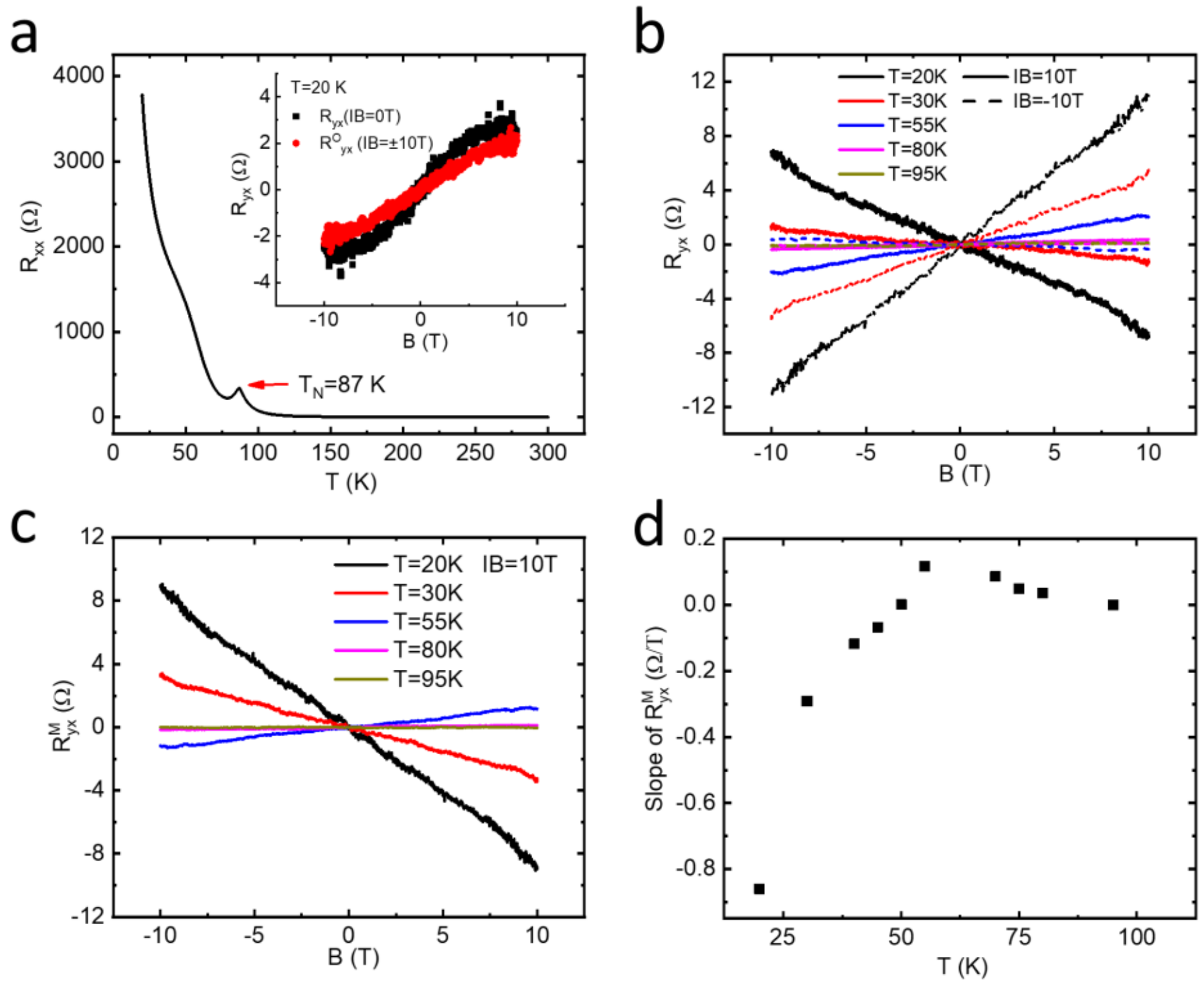

**Figure 5** Temperature dependence of unconventional component measured in sample N1 (second time). (a) Temperature dependence of the resistance of sample N1 (second time), showing a Néel temperature at 87.0 K which is slightly higher than the one measured for the first time. The inset shows $R_{yx}$ measured at 20 K after zero-field cooling from 100 K and $R_{yx}^{O}$ extracted from $R_{yx}$ measured at 20 K after field cooling from 100 K under IB=10/-10 T. (b) $R_{yx}$ measured at some typical temperatures after field cooling from 100 K under IB=10 T (solid lines) and -10 T (dashed lines). (c) $R_{yx}^{M}$ under IB=10 T at some typical temperatures extracted from the $R_{yx}$ results in (b). (d) Temperature dependence of the low-field slope of $R_{yx}^{M}$.